\documentclass[pre,preprint,amssymb,showkeys,superscriptaddress,notitlepage]{revtex4-1}

\pdfoutput=1

\usepackage{graphicx}
\usepackage{dcolumn}
\usepackage{amsmath}    
\usepackage{amssymb}
\usepackage{bm} 
\usepackage{hyperref}
\usepackage{latexsym}
\usepackage{verbatim}
\usepackage{color}
\usepackage[caption=false]{subfig}

\def\beq{\begin{equation}}
\def\eeq{\end{equation}}

\def\beq{\begin{equation}}                           
\def\eeq{\end{equation}}                           
\def\bea{\begin{eqnarray}}                           
\def\eea{\end{eqnarray}}        

\preprint{}

\begin{document}

\preprint{}

\title{Collection  of  polar self-propelled particles with  a modified alignment interaction}
\author{Sudipta Pattanayak}
\email{pattanayak.sudipta@gmail.com}
\affiliation{S N Bose National Centre for Basic Sciences, J D Block, Sector III, Salt Lake City, Kolkata 700106}
\author{Shradha Mishra}
\email[]{smishra.phy@itbhu.ac.in}
\affiliation{Department of Physics, Indian Institute of Technology (BHU), Varanasi, India 221005}

\date{\today}

\begin{abstract}
{We study the disorder-to-order transition in a collection of polar 
self-propelled  particles interacting through a distance dependent 
alignment interaction. Strength of the interaction, $a^{d}$ ($0<a<1$) 
decays with metric distance $d$ between particle pair, and 
the interaction is short range. At $a = 1.0$, our model 
reduces to the famous Vicsek model. For all ${\it a}>0$, the system shows a transition 
from a disordered to an ordered state as a function of 
noise strength. We calculate the critical 
noise strength, $\eta_c(a)$ for different $a$ and compare it with 
the mean-field result. Nature of the disorder-to-order transition 
continuously changes from discontinuous to continuous with decreasing $a$. 
We numerically  estimate tri-critical point $a_{TCP}$ at which the nature of 
 transition changes from discontinuous to continuous. 
The density phase separation is large for ${\it a}$ close to unity, 
and it decays with  decreasing $a$. We also write the coarse-grained 
hydrodynamic equations of motion for general ${\it a}$, and find that the 
homogeneous ordered state is unstable to small perturbation as
${\it a}$ approaches to $1$. The instability in the  homogeneous ordered 
state  is consistent with the large density phase separation  
for ${\it a}$ close to unity. }

\end{abstract}
\maketitle
\section{Introduction \label{introduction}}
Flocking \cite{bacterialcolonies, insectswarms, birdflocks, fishschools}, the collective and 
coherent motion of large number of organisms, is one of the most familiar and ubiquitous 
biological phenomena. In the last one decade, there have been an increasing interest in the  
rich behaviors of these systems that are different from their equilibrium counterparts 
\cite{sriramrev3, sriramrev2, sriramrev1}. One of the key features of these flocks is that 
there is a transition from a disordered state to a long ranged ordered state in two-dimensions with the variation 
of system parameters (e.g., density, noise strength)  \cite{vicsek1995, chate2007, chate2008}. 
The study of the phase transition in these systems is an active area of research, even after many years 
since the introduction of the celebrated model by Vicsek et. al. \cite{vicsek1995}. Many studies have 
been performed with different variants of metric distance model \cite{vicsektricritical} 
and topological distance model \cite{chatetopo, chatetopo1, biplab}. In the Novel work of Vicsek, 
it is observed that the disordered to ordered state  transition is continuous \cite{vicsek1995}, 
but later other studies \cite{chate2007, chate2008} confirmed that the transition is discontinuous. Some studies on the topological distance model claim the transition to be discontinuous \cite{biplab}, whereas other studies \cite{chatetopo, chatetopo1} find it continuous.
Therefore, the nature of the transition of polar flock is still a matter of debate.\\

In our present work we ask the question, whether the
nature of transition in polar flock can be tuned by tuning certain system parameters. And how do the characteristics of system change for the two types of transitions (discontinuous / continuous) ? 
To answer this, we introduce a distance dependent parameter $a$ such that 
the strength of interaction decays with distance. For 
$a=1$, the interaction is same as that in the Vicsek model. For all non-zero distance dependent parameter ($a > 0$), the system is in a disordered state at small density 
and high noise strength, and in an ordered state at high density and low noise strength. 
We calculate the critical noise strength $\eta_c(a)$ for different $a$ and compare 
it with the  mean-field result. The nature of the disorder to order transition continuously changes from 
discontinuous to continuous with decreasing $a$. We estimate the tri-critical point  
in the noise strength $\eta$ and $a$ plane, where the nature of the transition changes from discontinuous to continuous. We also calculate the density phase separation in the system. 
The density phase separation order parameter is large for $a$ close to unity, 
and it monotonically decays with decreasing $a$ . Linear stability analysis of the homogeneous ordered 
state shows an instability as $a$ approaches to $1$, which is consistent with large density phase separation for $a \simeq 1$. \\  

This article is organised as follows. In  section \ref{Model}, we introduce the microscopic 
rule based model for distance dependent interaction. The results of numerical simulation 
are given in section \ref{Numerical study}. In section \ref{hydrodynamics}, we write the 
coarse-grained hydrodynamic equation of motion, calculate the mean field estimate of critical 
$\eta_c(a)$, and discuss the results of linear stability analysis. Finally in section \ref{discussion}, 
we discuss our results and future prospect of our study. Appendix \ref{App:AppendixA} is at the end, that contains the detailed calculation of the linear stability analysis.

\section{Model \label{Model}}
We study a collection of polar self-propelled  particles on a two-dimensional substrate. 
The particles interact through a short range {\it alignment} interaction, which decays with the metric distance. \\ 
Each particle is defined by its position ${\bf r}_i(t)$ and orientation $\theta_i(t)$ or 
unit direction vector ${\bf n}_i(t)=[\cos\theta_i(t), \sin\theta_i(t)]$. Dynamics of the 
particles are given by two update equations. One for the position and other  for the orientation. 
Self-propulsion is introduced as a motion towards its orientation with a  
fixed step size($v_{0}$ in unit time). Hence, the position update equation of the particles
\begin{equation}
{\bf r}_{i}(t+1)={\bf r}_{i}(t)+v_{0}{\bf n_{i}},
\label{eqn1}
\end{equation} 
and the orientation update equation with a distance dependent short range alignment interaction
\begin{equation}
{\bf n_{i}} (t+1)=\frac{\sum _{j\in R_{0}} {\bf n_{j}}(t)a^{d}+N_{i}(t)\eta {\bf \zeta}_{i}}{W_{i}(t)}
\label{eqn2}
\end{equation} 
where the sum is over all the particles within the interaction radius ($R_{0}$) of the $i^{th}$ particle, i.e., 
$\vert {\bf r}_{j}(t)-{\bf r}_{i}(t)\vert < R_0(=1)$. $N_i(t)$ is the number of particles 
within the interaction radius of the $i^{th}$ particle at time t, and $d$ is the metric distance between a pair of particles 
$(i,j)$. $W_{i}(t)$ is the normalisation factor. The strength of the noise $\eta$ is varied between  zero to $1$, and ${\bf \zeta}_i(t)$ is a random unit vector. Note that this model reduces to the celebrated Vicsek model for $a = 1.0$. \\

\begin{figure}[ht]
\centering
\includegraphics[width=0.78\linewidth]{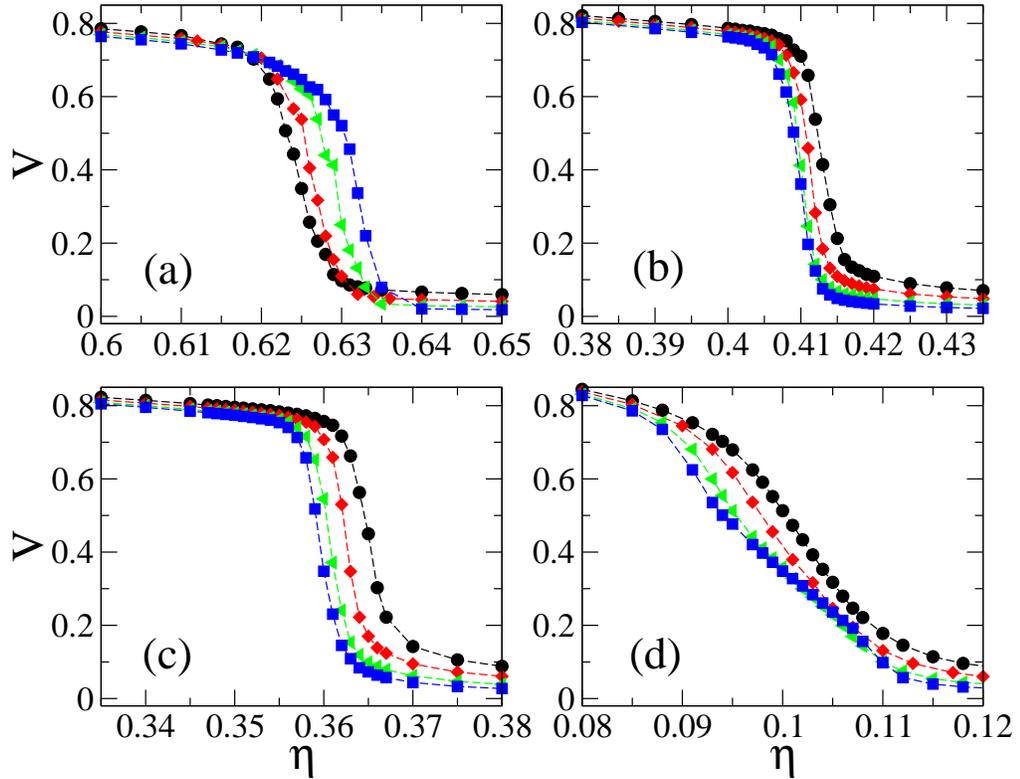}
\caption{[Color online] Plot of the global velocity V vs. the noise strength $\eta$ for four different distance dependent parameters $a$. Fig. (a-d) are for $a = 1.0, 0.5, 0.4, 0.01$ respectively. In Fig. (d), the variation of $V$ is clearly continuous for all system sizes, and there is no crossover. The variation of $V$ changes as we increase $a$, and there is a crossover for $a = 1.0$. Plot of the V for four different system sizes ( N = 1000, 2000, 5000, 10000) are shown by black  $\bullet$, red $\blacklozenge$ , green $\blacktriangle$ and blue $\blacksquare$ respectively.}
\label{fig:fig3}
\end{figure}

\begin{figure}[ht]
\centering
\includegraphics[width=0.78\linewidth]{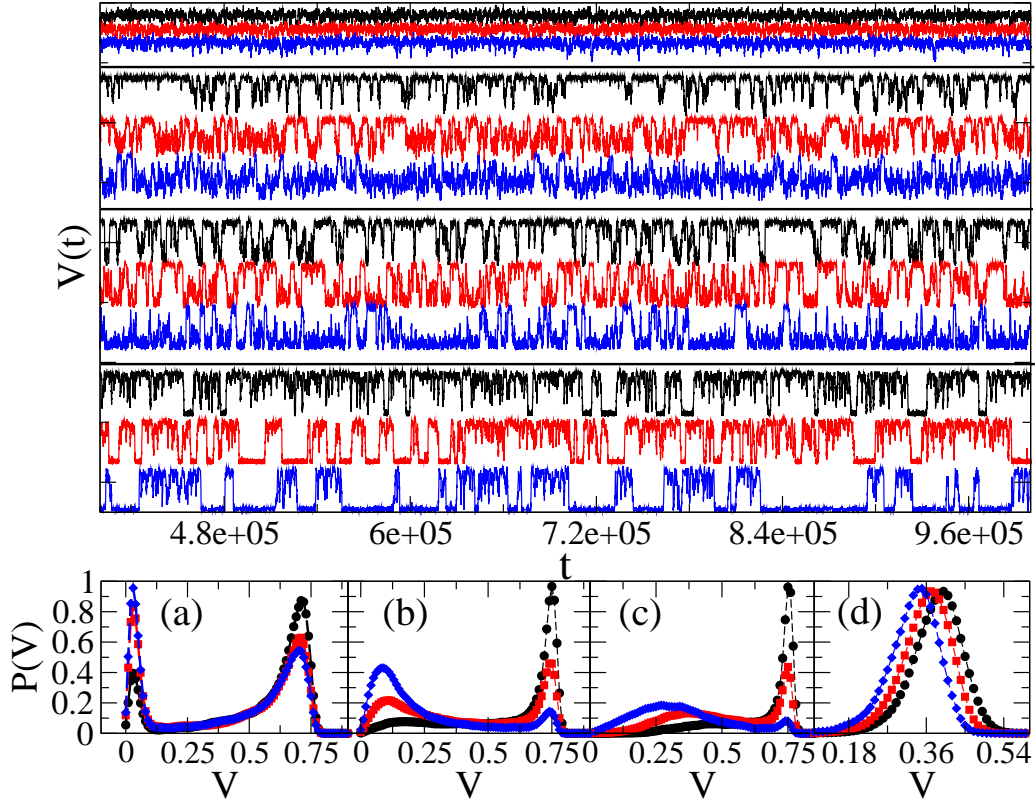}
\caption{[Color online] {\it Upper panel} : Plot of the time series of the global velocity  V for four different $a = (0.01, 0.4, 0.5, 1.0)$, from top to bottom. The time series of the V are plotted for three different noise strengths $\eta_1(a)$(black) $< \eta_2(a)$(red) $< \eta_3(a)$(blue) close to the critical noise strength $\eta_c$ for each $a$.  For $a = 0.01$ the time-series of the $V(t)$ is shown for $\eta_1 = 0.099$(black), $\eta_2 = 0.100$(red) and $\eta_3 = 0.101$(blue). Similarly $\eta_1 < \eta_2 < \eta_3$ for  $a = 0.4, 0.5$ and $1.0$ are $(0.358, 0.359, 0.360)$, $(0.409, 0.410, 0.411)$ and $(0.627, 0.628, 0.629)$ respectively.  There is a clear switching behavior in the global velocity variation for $a = 1.0$, and it vanishes as we decrease $a$. Time-series are shifted on the vertical axis for clarity. 
{\it Lower panel} : We plot the probability distribution function (PDF) of the global velocity $P(V)$ for four different $a = (1.0, 0.5, 0.4, 0.01)$ in Fig. (a - d) respectively. We consider three different $\eta$ for each $a$, same as in upper panel. In Fig. (a) plot of $P(V)$ is clearly bimodal, and as we decrease $a$ it becomes to uni-modal in Fig. (d). All the plots are for  $N = 5000$.}
\label{fig:fig6}
\end{figure}

\begin{figure}[ht]
\centering
\includegraphics[width=0.78\linewidth]{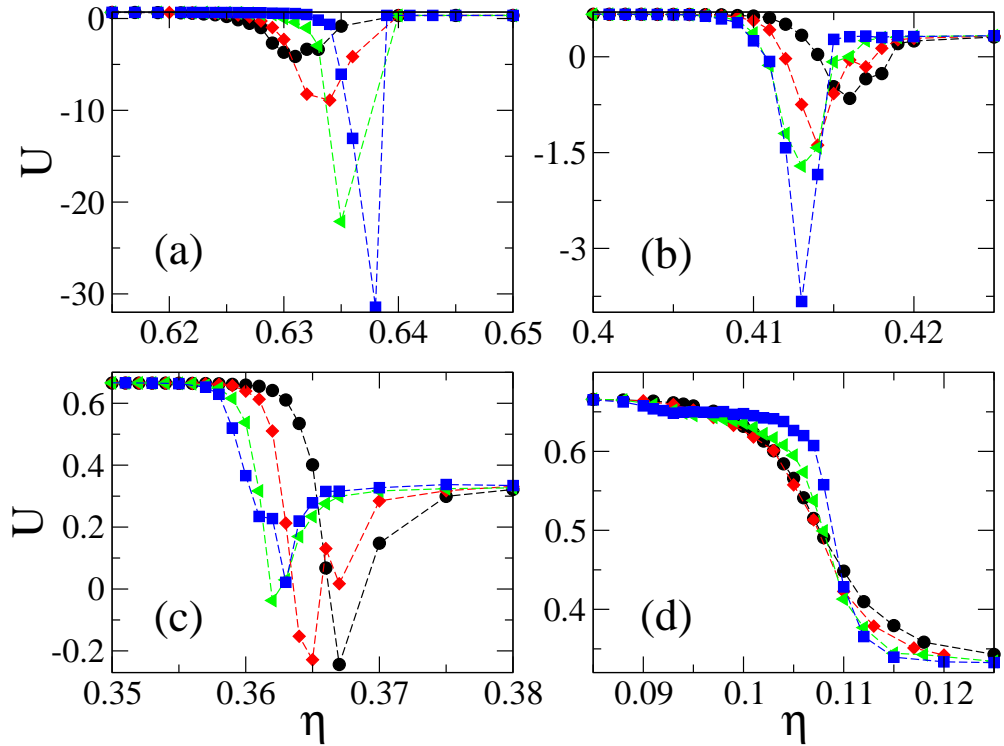}
\caption{[Color online] Plot of the Binder cumulant U vs. the noise strength $\eta$ for four different distance dependent parameter $a$. Fig. (a-d) are for $a = 1.0, 0.5, 0.4, 0.01$ respectively. $U$ varies discontinuously from  $1/3$ (disordered state) to $2/3$ (ordered state) in Fig. (a), and it goes continuously from  $1/3$ to $2/3$  in Fig. (d). Discontinuity in the variation of $U$  increases with system size for $a \gtrsim 0.4$, and it decreases for $a \lesssim 0.4$. Symbols have the same meaning as in Fig. \ref{fig:fig3}.}
\label{fig:fig5}
\end{figure}

\begin{figure}[ht]
\centering
\includegraphics[width=0.78\linewidth]{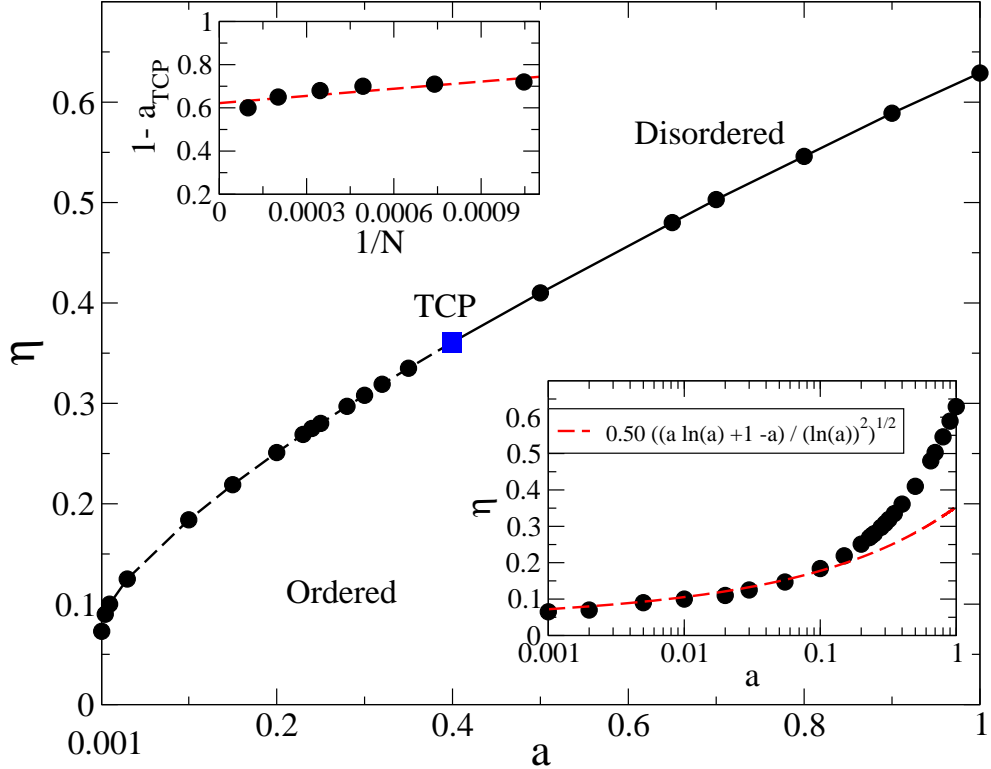}
\caption{{\it Main} : Schematic phase diagram of the disorder-to-order transition in 
noise strength $\eta$ and distance dependent parameter $a$ ($\eta,a$) plane. 
For all $a > 0$ there is a phase transition from a disordered to an ordered phase with decreasing $\eta$ across the critical noise strength line. Dashed line indicates the nature of the transition is continuous, whereas solid line indicates the discontinuous transition. The nature of transition changes from discontinuous to continuous at a tri-critical point $a_{TCP}$(square). {\it Lower inset}: we compare with the mean-field calculation of the critical noise strength $\eta_c$  for different $a$ with our numerical data. Mean field results fit well with numerical data for small values of $a$. In upper inset : plot of  $1-a_{TCP}$ vs. $1/N$ shows the variation of TCP with system size. We find $a_{TCP}$ converges to $a \approx 0.39$ for $N \rightarrow \infty$(thermodynamic limit).}
\label{fig:fig7}
\end{figure}

\begin{figure}[ht]
\centering
\includegraphics[width=0.5\linewidth]{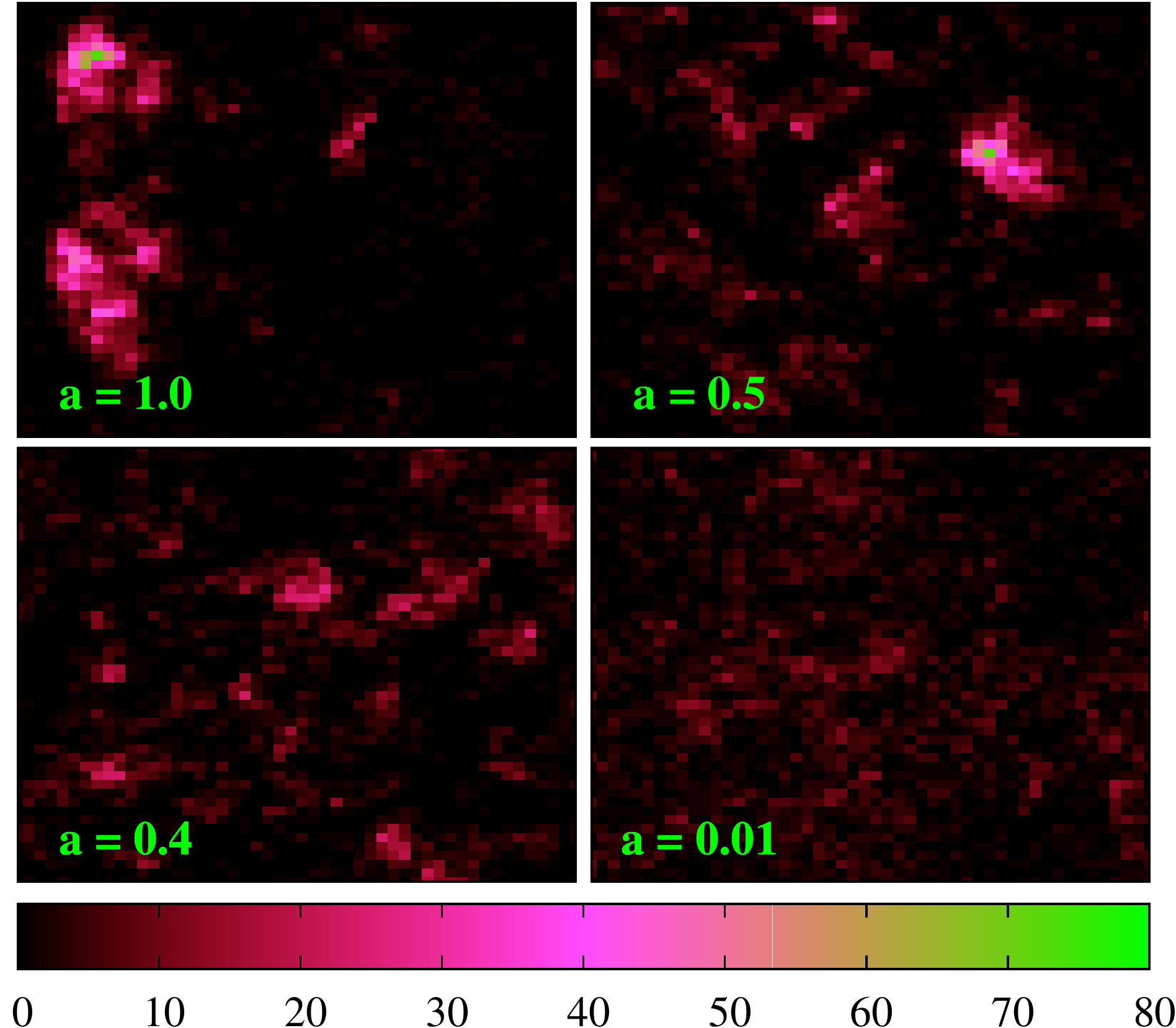}
\caption{[Color online] Plot of real space snapshots of the particle density distribution for four different $a(1.0, 0.5, 0.4, 0.01)$. {\it Upper panel}: Plot of the particle density distribution for $a = 1.0$ and $a = 0.5$ from left to right respectively. {\it Lower panel}: Plot of the particle density distribution for $a = 0.4$ and $a = 0.01$ in the same order. Color bar shows the number of particles in a unit sized sub-cell.}
\label{fig:fig2}
\end{figure}

\begin{figure}[ht]
\centering
\includegraphics[width=0.78\linewidth]{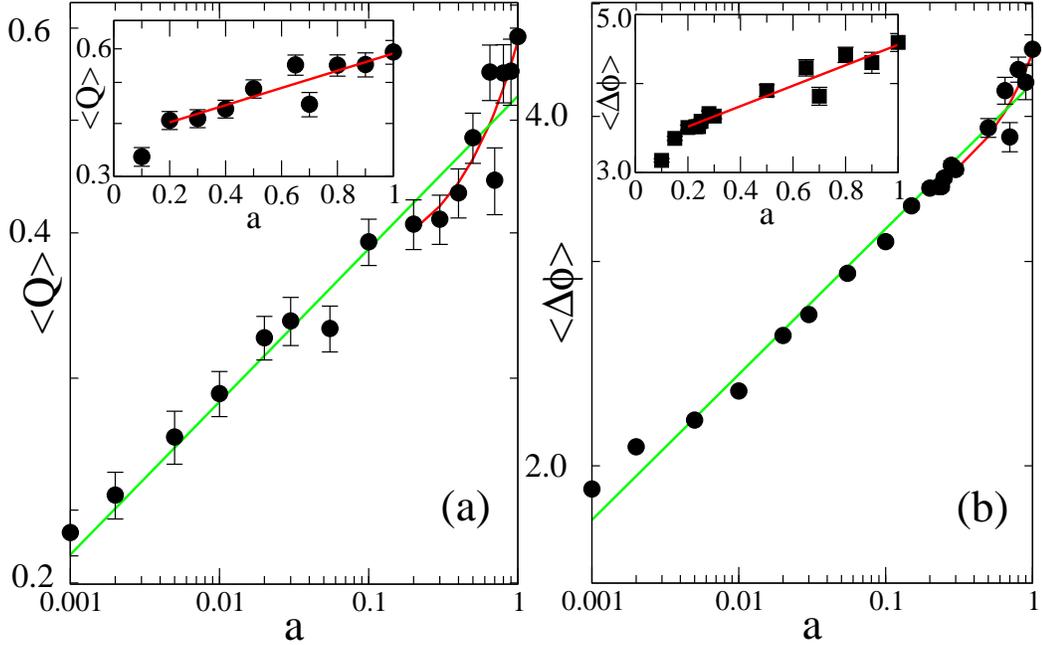}
\caption{ Plot of the average density phase separation order parameter $<Q>$ vs. $a$, and the average standard deviation in particle number in a unit cell $<\Delta \phi>$ vs. $a$ are shown in Fig. (a) and (b) respectively in log-log scale. 
 $<Q>$ and $<\Delta \phi>$ decay exponentially from $a = 1.0$ to $a \approx 0.2$. Both show similar power law decay with the exponent $0.13$, for small values of $a$. In the insets of Fig (a) and (b), we show the exponential decay of the $<Q>$ ($ \sim e^{0.46a}$) and $<\Delta \phi>$ ($ \sim e^{0.33a}$) in semi-log scale.}
\label{fig:fig8}
\end{figure}

\begin{figure}[ht]
\centering
\includegraphics[width=0.78\linewidth]{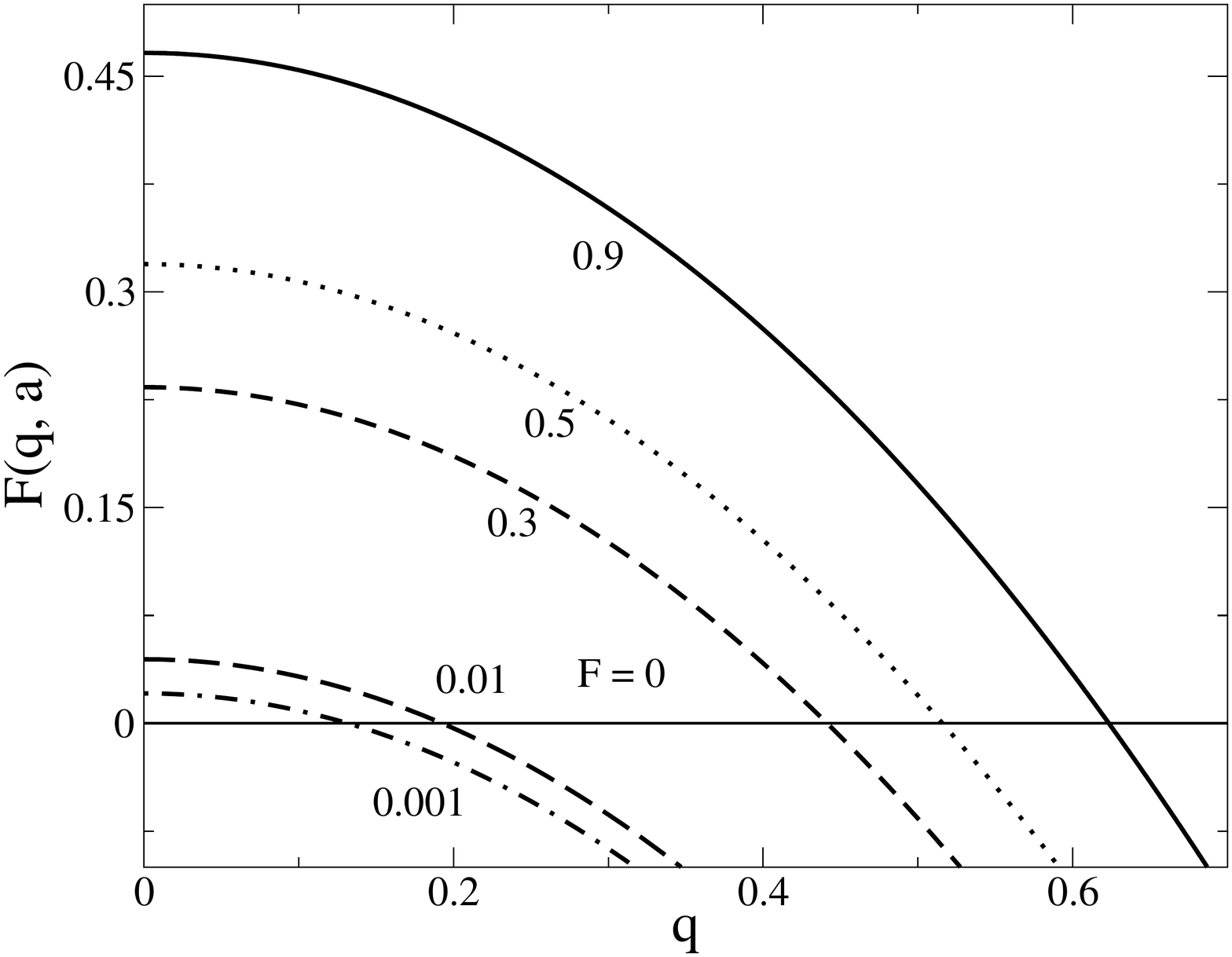}
\caption{Plot of  $F(q,a)= \frac{(a \ln(a) +1 -a)}{(\ln(a))^2} - \frac{D_Vq^{2}}{2}[(\frac{\lambda}{v_0}-1)+
\sqrt{(\frac{\lambda}{v_0}-1)^{2}+\frac{1}{2v_0}} ]$ vs. wave vector $q$. 
For  $v_0=0.5$, $D_v=1.0$, $\lambda=1.0$, $\alpha_0=1.0$.  
$F(q,a)$ becomes +ve for small  $q$, which suggests that hydrodynamic mode becomes unstable 
at smaller wave vector. Region of instability continuously increases with increasing $a$.}
\label{fig:fig9}
\end{figure}

\section{Numerical Study \label{Numerical study}} 
We numerically simulate the microscopic model introduced by Eqs.\ref{eqn1} and \ref{eqn2} 
for different distance dependent parameter $a$. 
For $a=1$, the particle interacts with the same strength with all the particles inside its interaction radius (Vicsek's model \cite{vicsek1995}). 
As we decrease $a$, interaction strength decays with distance. $a$ is varied from $1.0$ to small value $0.001$. For $a = 0.0$ the particles are non-interacting. Speed of  the particles is fixed to $v_0=0.5$. We start with random orientation and 
homogeneously distributed particles on a $2-$dimensional substrate of size $L \times L$ 
with periodic boundary conditions. For all the simulations, we keep mean density  $\rho_0=\frac{N}{L^2}=2.0$. 
Number of particles were varied from $N=1000$ to $10000$.  
We start from a random state and each particle is updated using Eqs. \ref{eqn1} and \ref{eqn2}. 
One simulation step is counted after sequential update of all the particles. 
All the measurements are performed after $10^5$ simulation steps, and a total of $10^{6}$ steps are used in simulations. 

\subsection{Disorder-to-order transition \label{disorder}}
First we study the disorder-to-order transition in the system for different $a$. Ordering in the system is characterised by the global velocity,
\begin{equation}
V=\vert\frac{1}{N} \sum _{i=1} ^N {\bf{n}}_{i}(t)\vert.
\label{eqn9}
\end{equation}
In the ordered state, i.e., when large number of particles are oriented in the same direction, then
 $V$ is close to 1,  and it is close to zero for a random disordered state. In Fig. \ref{fig:fig3} (a-d) we have shown the variation of $V$ with the noise strength $\eta$ for four different $a ( = 1.0, 0.5, 0.4, 0.01)$ respectively. For $a = 1$, on increasing $N$, the variation of $V$ shows a crossover behaviors. This kind of crossover is a common feature of first order transition \cite{chate2008, biplab}. Whereas for $a = 0.01$, $V$ varies continuously, and the transition is second order. The variation of $V$ in the intermediate region of $a$, changes smoothly from one type to another. We also estimate the critical $\eta_{c}(a)$ for different $a$ values, and it decreases with $a$, provided other parameters ({\it viz} mean density $\rho_0$,  speed $v_0$) are kept fixed.  \\ 

Now to characterize the nature of the transition with the variation of $a$, we plot the time series of the global velocity $V(t)$ for 
four different $a (= 0.01, 0.4, 0.5, 1.0)$, from top to bottom in the upper panel of Fig. \ref{fig:fig6}. We choose three different $\eta$ ($\eta_{1}(a)$(black) $< \eta_{2}(a)$(red) $< \eta_{3}(a)$(blue)) for each $a$ close to the critical noise strength $\eta_c(a)$. 
For $a = 1$, we choose $\eta_{1} = 0.627$, $\eta_{2} = 0.628$ and $\eta_{3} = 0.629$, and plotted the time-series of $V$. 
$V(t)$ shows switching  behaviour, and it alternates between two finite values of $V$. 
$V(t)$ keeps switching throughout the simulation time. 
At smaller $a=0.5$ ($\eta_{1} = 0.409, \eta_{2} = 0.410, \eta_{3} = 0.411$) we again find switching behaviour, 
but the difference between two finite values of $V$ decreases. 
Switching behaviour further reduces for $a=0.4$  $(\eta_{1} = 0.358, \eta_{2} = 0.359, \eta_{3} = 0.360)$.
For  small $a = 0.01$ ($\eta_{1} = 0.099, \eta_{2} = 0.100, \eta_{3} = 0.101$) $V(t)$ shows fluctuations, 
but there is no switching behaviour. We further calculate probability distribution $P(V)$ of the global velocity for the same set of $a$ and $\eta$ values as used for the time series plots. As shown in Fig. \ref{fig:fig6}(a), $P(V)$ is bimodal for $a = 1.0$, i.e., there are two distinct peaks for $P(V)$. Two finite values of $V$ corresponds to two states of the system.
 Two peaks come closer with decreasing $a$, and for small $a (=0.01)$, $P(V)$ shows only one broad peak in Fig. \ref{fig:fig6}(d). The bimodal distribution of the $V$ confirms that the transition is discontinuous for $a \simeq 1$. \\

To further characterise the nature of the transition, 
we calculate the fourth order cumulant or the Binder cumulant, i.e.,
\begin{equation}
U=1-\frac{<V^{4}>}{3<V^{2}>^{2}}
\label{eqn10}
\end{equation}
$U(\eta)$ vs. $\eta$ plot is shown in Fig. \ref{fig:fig5}. It  shows strong discontinuity 
from $U = 1/3$ (for disordered state) to $U = 2/3$ (for ordered state) as we approach critical 
$\eta_c(a)$ for $a=1$ in Fig. \ref{fig:fig5} (a), and discontinuity decreases with $a$. 
It smoothly goes from a disordered state ($U = 1/3$) to an ordered state ($U = 2/3$) for $a=0.01$ in Fig. \ref{fig:fig5} (d). For $a \gtrsim 0.4$, $U$ vs. $\eta$ plot shows strong discontinuity at large $N$, but for $a \lesssim 0.4$ it becomes continuous.\\

Therefore, The nature of the transition continuously changes from discontinuous   
to continuous  on decreasing $a$. The critical noise strength $\eta_c(a)$ also decreases 
with decreasing $a$. We plot $\eta_c(a)$ vs. $a$ in the of Fig. \ref{fig:fig7}. 
The solid  line indicates the nature of the disorder-to-order transition is discontinuous, and the dashed line indicates the continuous transition.  
The value of $a$ at which the above transition changes from discontinuous to continuous one, 
we call it as tri-critical-point (TCP) $a_{TCP}$. For  $a >  a_{TCP}$ the transition is discontinuous, 
and for $a < a_{TCP}$ it is continuous. TCP shows a small 
dependence on $N$ for any fixed $v_0$ and $\rho_0$. We define the TCP for any system size as the point where the Binder cumulant $U$ starts to show discontinuous variation. In the upper inset of Fig. \ref{fig:fig7}, we plot $1-a_{TCP}$ vs. $1/N$, 
and extrapolate the  TCP for $N \rightarrow \infty$ or  $1/N$ $\rightarrow$ zero. 
As $1/N$ approaches to zero, $1-a_{TCP} \approx 0.61$. Hence the $a_{TCP}$ is $\approx 0.39$. Hence, the extrapolated value of $a_{TCP}$ matches well with the $a_{TCP}$ in phase diagram, which is marked 
as blue square in Fig. \ref{fig:fig7}.  In the lower inset of Fig. \ref{fig:fig7}, we 
plot the critical $\eta_c(a)$ vs. $a$ on semi-log scale and compare the 
results with the mean field result in Eq. \ref{eta1}. 
Mean field approximation is good when density distribution is homogeneous. In such limit, 
density at each point is close to the mean density of the system. As shown in Fig. \ref{fig:fig2} density distribution 
becomes more and more inhomogeneous as we increase $a$. Hence, for the
small $a$ values numerical estimate of $\eta_c(a)$ should  be more close to MF. We show in lower inset of Fig. \ref{fig:fig7} the 
 numerical $\eta_c(a)$ matches very well with MF for small $a < 0.1$.

\subsection{Density phase separation \label{denps}}
The density distribution of particles also changes as we vary $a$. 
Density fluctuation plays an important role in 
determining the nature of the transition in polar flock \cite{chate2008, solon, solon1, chate2004, shradhaprl, das2012, aditi}. 
In Fig. \ref{fig:fig2} we show the real space snapshot of particle density for 
different $a (= 1, 0.5, 0.4$ and $0.01)$ close to  critical noise strength $\eta_c(a)$. 
Clusters are small and homogeneously distributed for small $a$, but as $a$ approaches to $1$ we find large, dense and anisotropic clusters.  
We quantify the density distribution by calculating the density phase separation order parameter in Fourier space defined as,
\begin{equation}
Q({\bf k})=\mid \frac{1}{L} \sum _{i,j=1} ^{L} e^{i{\bf k}\cdot {\bf r}} \rho(i,j) \mid
\label{eqn11}
\end{equation}
where ${\bf k}=\frac{2 \pi (m, n)}{L}$ is a two dimensional wave vector and $m,n$ = $0$, $1$, $2$ ...., $L-1$ . The reference frame is chosen so that the orthogonal axes $(1,0)$ and $(0,1)$ are along the boundary of the substrate, and  $(1,1)$ represents diagonal direction. We calculate the first non-zero value of $Q({\bf k})$ in all three directions $Q(1,0)$, $Q(0,1)$ and $Q(1,1)$. The average density phase separation order parameter $<Q>$ is $(Q(1,0)+Q(0,1)+Q(1,1)) / 3$. \\

We also characterize the density phase separation using the  standard deviation in particle number $\Delta \phi$ in a unit size sub-cell. It is defined as 
\begin{equation}
\Delta \phi =\sqrt{\frac{1}{N_{c}}\sum_{j=1}^{N_{c}}(\phi_{j})^{2}-(\frac{1}{N_{c}}\sum_{j=1}^{N_{c}}\phi_{j})^{2}}
\label{eqn12}
\end{equation}
where $\phi_{j}$ is the number of particles in the $j^{th}$ sub-cell. To calculate $\Delta \phi$ we first divide the whole system into $N_c( = L^2)$ unit sized sub-cells, then calculate the number of particles in each sub-cell, and from there we calculate the standard deviation in particle distribution. $Q(t)$ and $\Delta \phi(t)$
are calculated at different times in the steady state, and then average
over a large time to obtain $<Q>$ and $<\Delta \phi>$ respectively. Plots of $<Q>$ and $<\Delta \phi>$ vs. $a$ on  log-log scale 
are shown on Fig. \ref{fig:fig8} (a) and (b)  respectively. For $a\simeq 1$ both  $<Q>$ and $\Delta \phi$ are large; however , as we decrease $a$, they   
decay monotonically. For $a$ close to unity both $<Q>$ and $<\Delta \phi>$ show fast decay ({\it exponential}), and 
for smaller $a$ they decay algebraically
with $a$. In the insets of Fig. (a) and (b), we show the exponential decay of  the density phase separation order parameter $<Q>$ ($ \sim e^{0.46a}$), and the standard deviation in particle distribution $<\Delta \phi>$ ($ \sim e^{0.33a}$) 
for $a \approx 1$. We find that for $a\approx 1$, the density phase separation is high, and the nature of the disorder-to-order transition is also first order. Hence, the change in the nature of both the disorder-to-order transition and the density phase separation shows
 variation on decreasing $a$.\\

\section{Hydrodynamic equations of motion}\label{hydrodynamics}
We estimate the $\eta_c(a)$ and also study the linear stability of homogeneous ordered state with varying $a$. 
The coarse-grained hydrodynamic variables are 
coarse-grained density $\rho(r,t)$ and velocity $V(r,t)$ and they are defined as,
\begin{equation}
\rho({\bf r},t)=\sum_{i=1}^{N}\delta({\bf r}-{\bf r}_{i}(t))
\label{eqn3}
\end{equation}
\begin{equation}
{\bf V}({\bf r},t)=\dfrac{\sum_{i=1}^{N}v_0{\bf n}_{i}(t)\delta({\bf r}-{\bf r}_{i}(t))}{\rho({\bf r},t)}
\label{eqn4}
\end{equation}
We can write the coupled hydrodynamic equations of motion for density and velocity as obtained in Toner and Tu \cite{tonertu}
\begin{equation}
\partial_{t}\rho=-v_{0}{\nabla}.(\rho {\bf V} )
\label{eqn5}
\end{equation}
and for velocity
\begin{equation}
\partial _{t} {{\bf V}} =  \alpha (\rho, \eta, a){\bf V} -\beta (\mid V \mid)^2 {\bf V} - \frac{v _{1}}{2\rho _{0}} {\nabla} \rho  + D_{V} \nabla^{2}{\bf V}  -\lambda_{1} ({\bf V}.{\nabla}) {\bf V}-\lambda_{2}({\bf \nabla}. {\bf V}){\bf V} -\lambda_{3} \nabla (\mid V\mid ^{2})
\label{eqn6}
\end{equation} 
For our distance dependent model  we have  introduced an additional general $a$ dependence to alignment 
parameter $\alpha(\rho, \eta, a)$ in the velocity equation \ref{eqn6}. In \cite{tonertu} $\alpha$ is
treated as a constant. But in general $\alpha$ is a function of  microscopic parameters (e.g. density, noise strength etc.) when derived from microscopic model. 
For $a=1$, our model reduces to  the  Vicsek's model, and $\alpha=\alpha_0(\rho-\rho_c)$.
$\rho_c$ in general depends on system parameters ({\it viz}: noise strength, speed etc.)
On increasing   density
large noise is required to break the order or  $\rho_c$ increases with $\eta$. Using mean-field-like
argument it can be shown that $\rho_c \simeq \frac{\eta^2}{v_{0}^{2}}$ \cite{chate2008} or $\alpha=\alpha_0(\rho-4\eta^2)$.
$\alpha$ shows linear dependence on $\rho$ for $a=1$, when all the particles
within the coarse-grained radius interact with same strength. In general for 
$a<1$,  strength of interaction decays with distance. 
Again using  the mean-field limit when density inside the coarse-grained 
radius is homogeneous, following form of $\alpha$ is obtained 
\begin{equation}
\alpha(\rho, a, \eta) = \alpha_0 \left(\rho_0 [\frac{(a \ln(a) +1 -a)}{(\ln(a))^2}] -8 \eta^2\right)
\label{alpha}
\end{equation}
Hence $\alpha$ changes sign at critical $\eta_c$. 
\begin{equation}
\eta_c(a) = \sqrt{\frac{\rho_0}{8}} \sqrt{\frac{(a \ln(a) +1 -a)}{(\ln(a))^2}}
\label{eta}
\end{equation}
Which for mean density $\rho_0=2.0$ reduces to
\begin{equation}
\eta_c(a) = \frac{1}{2} \sqrt{\frac{(a \ln(a) +1 -a)}{(\ln(a))^2}}
\label{eta1}
\end{equation}
The  
 homogeneous solution for the disordered state is  $V_0=0$ (for $\eta > \eta_c$), and for the ordered state is
 $V_{0} = \sqrt{\frac{\alpha(\rho_0,a)}{\beta}}$ (for $\eta < \eta_c$). 
 
In Fig. \ref{fig:fig7} (lower inset) we plot the function $\eta_c(a)$ vs. $a$ as given in 
Eq.\ref{eta1} on semi-log scale and its comparison to numerically estimated 
$\eta_c(a)$. We find that the data matches very well
with numerical result for small $a$ limit. Deviation from the MF expression increases with increasing $a$ when
the density distribution becomes more inhomogeneous Fig. \ref{fig:fig2}.\\

Now we study  the linear stability analysis of Eqs. \ref{eqn5} and \ref{eqn6} 
about the homogeneous ordered state for general $a$. Detail steps of linear stability analysis are given  in the appendix \ref{App:AppendixA}. We find that for large
$a$ homogeneous ordered state is unstable with respect to small perturbation. The condition for the instability is obtained in Eq. \ref{eqn26}.
\begin{equation}
\alpha_{1}^{\prime}> \frac{D_Vq^{2}}{2}[(\frac{\lambda}{v_0}-1)+\sqrt{(\frac{\lambda}{v_0}-1)^{2}+\frac{1}{2v_0}} ]
\label{eqn51}
\end{equation}
where $\alpha'_{1}(\rho_0) = \frac{d \alpha(\rho)}{d \rho}|_{\rho=\rho_0}$ = $\alpha_0 \left([\frac{(a \ln(a) +1 -a)}{(\ln(a))^2}] \right)$. Hence, using
the expression for $\alpha$ from Eq. \ref{alpha} we get condition for instability of the hydrodynamic mode,
\begin{equation}
\begin{aligned}
\alpha_0 \frac{(a \ln(a) +1 -a)}{(\ln(a))^2}  - \frac{D_Vq^{2}}{2}[(\frac{\lambda}{v_0}-1)+ \sqrt{(\frac{\lambda}{v_0}-1)^{2}+\frac{1}{2v_0}} ] > 0
\label{eqn41}
\end{aligned}
\end{equation}
We plot $F(q, a)=\alpha_0 \frac{(a \ln(a) +1 -a)}{(\ln(a))^2} - 
\frac{D_Vq^{2}}{2}[(\frac{\lambda}{v_0}-1)+\sqrt{(\frac{\lambda}{v_0}-1)^{2}+
\frac{1}{2v_0}} ]$ vs. $a$ in Fig. \ref{fig:fig9}, and find that the instability of the hydrodynamic mode increases with $a$.
Unstable homogeneous state  for  $a \approx 1$ is consistent with the large density phase separation obtained in numerical simulation. System  shows first order disorder-to-order transition for large $a$. As we decrease $a$ the nature of the transition changes continuously, and also the density phase separation decays. 

\section{Discussion \label{discussion}}
We introduce a variant of the Vicsek model \cite{vicsek1995} for the collection of polar self propelled 
particles with a modified alignment interaction.
Our model is similar  to  the celebrated Vicsek model  for  $a=1.0$.
Numerical simulations reveal that for all $a > 0$, the
system shows a transition from a disordered (global velocity $V \approx 0$)
to an ordered state (finite global velocity) 
on decreasing noise strength $\eta$, and the critical noise 
strength $\eta_c(a)$ also decreases with $a$. We find that
in a homogeneous system the disordered to ordered transition can be 
discontinuous or continuous depending on the distance dependent 
parameter ${\it a}$. The nature of 
the transition is characterized by 
calculating (a) the global velocity $V$,  (b) the fourth order variance in the global
velocity (Binder cumulant $U$), and  (c) the  probability distribution of 
the global velocity for different distance dependent parameter ${\it a}$.
 For the discontinuous transition, $U$ shows a strong discontinuity close to critical noise 
strength $\eta_c(a)$. The 
 variation of $V$ with time also shows switching between two states, and the probability
 distribution of the global velocity is bimodal for $a \approx 1$. However, for the continuous transition, $V$ 
continuously varies from large to small values and $U$ changes smoothly, and there is
 no switching behaviour in the global velocity time series, also the probability 
 distribution of the global velocity is uni-modal.    \\
We construct the phase diagram 
in the noise strength and the distance dependent parameter $(\eta , {\it a})$ plane.
The nature of the disorder-to-order transition is first order for $a \simeq 1$, 
and it changes  to continuous type with decreasing ${\it a}$, 
and at a tri-critical point the nature of the transition changes from discontinuous to continuous. 
Earlier studies of \cite{solon, solon1} find  that 
the disorder-to-order transition in polar flock can be mapped to  the liquid-gas 
 transition. In our study, we find that the density plays an important role and the large density inhomogeneity leads to the  
discontinuous transition in these systems.
  The effect of density is characterized by 
the phase separation order parameter $<Q>$ and the standard deviation in 
number of particles in unit sized sub-cells $<\Delta \phi>$ for different $a$. 
We find that the density phase separation is large for ${\it a} \simeq 1$, and 
 it decays with decreasing ${\it a}$. Hence, the discontinuous 
disorder-to-order transition and the large density phase 
separation are common for ${\it a}$ approaching to unity. \\
Our study concludes  that the nature of the 
disorder-to-order transition in collection of polar flock is 
not always necessarily first order, and 
it strongly depends on the interaction amongst the particles.
The study of \cite{vicsektricritical} shows that the transition from random  to  collective motion 
changes from continuous to discontinuous with decreasing restriction 
angle. The critical noise amplitude also decreases monotonically on decreasing the restriction
angle. In our model we propose a parameter ${\it a}$, which can also tune the nature of 
such transition. Our model would be useful to study the disorder-to-order 
 transition in biological and granular systems, 
where interaction between close-by neighbours is stronger than the interaction of particles with other neighbours.
     
\begin{acknowledgments}
{S. Pattanayak would like to thank Dr. Manoranjan Kumar for his kind cooperation and useful suggestions through out this work. S. Pattanayak would like to thank Department of Physics IIT (BHU), Varanasi for  kind hospitality. S. Mishra would like to thank DST for their partial financial support in this work.}
\end{acknowledgments} 

\appendix
\section{Linearised study of the broken symmetry state} \label{App:AppendixA}
The hydrodynamic equations Eq.\ref{eqn5} and \ref{eqn6} admit two homogeneous solutions: an isotropic state with ${\bf V}=0$ for $\rho < \rho_{c}$ and a homogeneous ordered state with ${\bf V}=V_{0}{\bf x}$ for $\rho > \rho_{c}$, where ${\bf x}$ is the direction of ordering. We are mainly interested in the symmetry broken phase. For $\alpha(\rho) > 0$ we can write the velocity field as ${\bf{V}}=(V_{o}+\delta V_{x}){\bf x}+\delta {\bf V}_{y}$, where ${\bf x }$ is the direction of broken symmetry and ${\bf y}$ is the perpendicular direction. $V_{0}{\bf x}=<{\bf{V}}>$ is the spontaneous average value of $\bf{V}$ in ordered phase. We choose $V_0=\sqrt{\frac{\alpha(\rho_{0},a)}{\beta}}$ and $\rho = \rho _{0}+\delta \rho$ where $\rho _{0}$ is coarse-grained density. Combining the fluctuations we can write in a vector format,
\begin{equation}
\delta X_{\alpha}({\bf r},t)=\left[ \begin{array}{c} \delta \rho \\ \delta V _{x} \\ \delta V_{y} \end{array} \right]
\label{eqn13}
\end{equation}
Now we introduce fluctuations in hydrodynamic equation for density and if we consider only linear terms then Eq.\ref{eqn5} will reduce to,
\begin{equation}
\partial_{t} \delta \rho + v_{0}V_{0}\partial_{x} \delta \rho  +v_{0}\rho_{0}\partial_{x} \delta {V_{x}}+v_{0}\rho_{0}\partial_{y} \delta {V_{y}}=0
\label{eqn14}
\end{equation} 
We consider the velocity fluctuation only in the direction of orientational ordering. So  $\delta {V_{y}}$ and $q_y$ is zero in our analysis. Now density Eq. \ref{eqn14} we can write as, 
\begin{equation}
\partial_{t} \delta \rho + v_{0}V_{0}\partial_{x} \delta \rho  +v_{0}\rho_{0}\partial_{x} \delta {V_{x}}=0
\label{eqn15}
\end{equation}
Similarly  we introduce fluctuations in velocity Eq. \ref{eqn6} and we are writing  velocity fluctuation equation for ordering direction. We also introduce functional density dependency in {\bf $\alpha(\rho)$}. We have done Taylor series expansion of $\alpha(\rho)$  in Eq.\ref{eqn6} at $\rho=\rho_0$, and consider upto first order derivative term of {\bf $\alpha(\rho)$}. Now velocity equation will reduces to,
\begin{equation}
\begin{aligned}
\partial _{t} {\delta {V _{x}}} = {} & \ (\alpha (\rho _{0})+ \alpha {_{1}^{\prime}} ({\rho _{0}})\delta \rho) (V_{0} + \delta V_{x}) -\beta (V_{0}^{2} +2V_{0} \delta V_{x})(V_{0}+ \delta V_{x})  -\frac{v _{1}}{2\rho _{0}} \partial _{x}  \delta \rho  \\ & + D_{V} \partial _{x} ^{2} \delta V _{x}+D_{V} \partial _{y} ^{2} \delta V _{x} -\lambda V _{0} \partial _{x} \delta V_{x} 
\label{eqn16}
\end{aligned}
\end{equation}
where $\alpha_{1}^{\prime}=\frac{\partial \alpha}{\partial \rho}\mid _{\rho _{0}}$ also $\lambda$ is combination of three $\lambda's(\lambda=\lambda_{1}+\lambda_{2}+2 \lambda_{3})$ terms. \\
Now considering no fluctuation along perpendicular direction of velocity field, equation along ordering direction(x-direction) reduces to,
\begin{equation}
\partial _{t} {\delta {V _{x}}}  + 2\alpha(\rho_0){\delta {V _{x}}} + \lambda V_0 \partial_x - D_V \partial_x^2{\delta {V _{x}}} - \alpha_1^\prime V_0 \delta \rho  + \frac{v_1}{2\rho_0} \partial_x \delta \rho = 0
\label{eqn17}
\end{equation}
Now we are introducing Fourier component, $\Delta Y(q,S) = \int dr \exp(i{\bf q.r}) \exp(St)dt$ in above two fluctuation equations \ref{eqn15}, \ref{eqn17} . Then we are writing the coefficient matrix for the coupled equations. Here we are writing $q_x = q$.
\begin{equation}
\begin{aligned}
\left[ \begin{array}{cc}  S + iv_{0}V_{0}q & iv_{0}\rho_{0}q \\ i \frac{v _{1}}{2 \rho _{0}} q -\alpha {_{1}^{\prime}}(\rho _{0}) V _{0} & S  +2\alpha +D _{V}q^{2} +i\lambda V _{0}q  \end{array} \right]  
\label{eqn18}
\end{aligned}
\end{equation}
Earlier study \cite{shradhapre, bertin} finds horizontal fluctuation or fluctuation in the direction of ordering is important when system is close to transition. Here important thing is that unlike isotropic problem $d>2$ there is no transverse mode, we always have just two longitudinal Gold-stone modes associated with $\delta \rho$ and $V_{x}$.
We get solution for hydrodynamic modes in symmetry broken state, 
\begin{equation}
S_{\pm}=-ic_{\pm}q-\epsilon_{\pm}
\label{eqn23}
\end{equation}
where the sound speeds, 
\begin{equation}
c_{\pm}=\frac{1}{2}(\lambda+v_{0})V_{0}\pm c_{2}
\label{eqn20}
\end{equation}
with
\begin{equation}
c_{2}=\frac{1}{2}\sqrt{(\lambda -v_{0})^{2}V_{0}^{2}+\frac{v_{0}v_{1}}{2}}
\label{eqn21}
\end{equation}
and the damping $\varepsilon _{\pm}$ in the Eq. \ref{eqn23} are $O({\bf q^{2}})$  and given by, 
\begin{equation}
\varepsilon _{\pm}=\pm \frac{c_{\pm}}{2c_{2}}[2\alpha+D_{V}q^{2}]\mp\frac{1}{2c_{2}}[2\alpha v_{0}V_{0}+v_{0}V_{0}\alpha _{1}^{\prime}+v_{0}V_{0}D_{V}q^{2}]
\label{eqn22}
\end{equation}
So real part of the modes are $-\epsilon_{\pm}$. Now we know the instability conditions are $1)$ If Re$[S_{\pm}]>0$ we will get homogeneous polarized state, which is unstable. $2)$ If Re$[S_{\pm}]<0$ we will get homogeneous polarized state, which is stable to small perturbation.
We know the expression for $\epsilon_{\pm}$,
\begin{equation}
\epsilon_{\pm}=\pm \frac{c_{\pm}}{2c_2}[D_{V}q^{2}+2\alpha] \mp \frac{1}{2c_2}[2\alpha v_0 V_0+v_0 V_0 \alpha_{1}^{\prime}+v_0 V_0 D_V q^{2}]
\label{eqn24}
\end{equation}
Close to transition point $\alpha \simeq 0$. So we can write,
\begin{equation}
\epsilon_{\pm}=\pm \frac{c_{\pm}}{2c_2}[D_{V}q^{2}] \mp \frac{1}{2c_2}[v_0V_0\alpha_{1}^{\prime}+v_0V_0D_Vq^{2}]
\label{eqn25}
\end{equation}
We have checked $Re[S_{-}]=-\epsilon_{-}<0$ always holds, so this mode is always stable. $Re[S_{+}]=-\epsilon_{+}>0$ for
\begin{equation}
\alpha_{1}^{\prime}> \frac{D_Vq^{2}}{2}[(\frac{\lambda}{v_0}-1)+\sqrt{(\frac{\lambda}{v_0}-1)^{2}+\frac{1}{2v_0}} ] ,
\label{eqn26}
\end{equation}
 and then this mode becomes unstable.




\end{document}